\documentclass[pdflatex,sn-mathphys,Numbered,iicol,a4paper]{sn-jnl} % double column layout
\usepackage{graphicx}%
\usepackage{multirow}%
\usepackage{amsmath,amssymb,amsfonts}%
\usepackage{amsthm}%
\usepackage{mathrsfs}%
\usepackage[title]{appendix}%
\usepackage{xcolor}%
\usepackage{textcomp}%
\usepackage{manyfoot}%
\usepackage{booktabs}%
\usepackage{algorithm}%
\usepackage{algorithmicx}%
\usepackage{algpseudocode}%
\usepackage{listings}%
\usepackage{breakurl}%

\begin{document}

\title[Resource-aware Research on Universe and Matter]{Resource-aware Research on Universe and Matter: Call-to-Action in Digital Transformation}

\author[1]{\fnm{Ben} \sur{Bruers}} % 
\author[2]{\fnm{Marilyn } \sur{Cruces}} % 0000-0001-6804-6513
\author[3]{\fnm{Markus} \sur{Demleitner}} % 
\author[4]{\fnm{Guenter} \sur{Duckeck}} % 0000-0002-7756-7801
\author[5]{\fnm{Michael} \sur{D\"uren}} % 0000-0002-6066-4744
\author[6]{\fnm{Niclas} \sur{Eich}} % 
\author[7]{\fnm{Torsten} \sur{Enßlin}} % 
\author[6]{\fnm{Johannes} \sur{Erdmann}} % 
\author*[6]{\fnm{Martin} \sur{Erdmann}}\email{erdmann@physik.rwth-aachen.de} % 
\author[6]{\fnm{Peter} \sur{Fackeldey}} % 
\author[8]{\fnm{Christian} \sur{Felder}} % 0000-0003-3746-213X
\author[6]{\fnm{Benjamin} \sur{Fischer}} % 0000-0002-3900-3482
\author[9]{\fnm{Stefan} \sur{Fr\"ose}} % 
\author[10]{\fnm{Stefan} \sur{Funk}} % 
\author[1]{\fnm{Martin} \sur{Gasthuber}} % 
\author[11]{\fnm{Andrew} \sur{Grimshaw}} % 
\author[9,12]{\fnm{Daniela} \sur{Hadasch}} % 
\author[8]{\fnm{Moritz} \sur{Hannemann}} % 
\author[2]{\fnm{Alexander} \sur{Kappes}} % 0000-0002-2000-3159
\author[13]{\fnm{Raphael} \sur{Kleinem\"uhl}} % 
\author[14]{\fnm{Oleksiy M.} \sur{Kozlov}} % 0000-0001-7394-2718
\author[4]{\fnm{Thomas} \sur{Kuhr}} % 0000-0001-6251-8049
\author[15]{\fnm{Michael} \sur{Lupberger}} % 
\author[13]{\fnm{Simon} \sur{Neuhaus}} % 
\author[1]{\fnm{Pardis} \sur{Niknejadi}} % 0000-0001-6685-674X
\author[16]{\fnm{Judith} \sur{Reindl}} % 
\author[17]{\fnm{Daniel} \sur{Schindler}} % 
\author[8]{\fnm{Astrid} \sur{Schneidewind}} % 
\author[18]{\fnm{Frank} \sur{Schreiber}} % 0000-0003-3659-6718
\author[19]{\fnm{Markus} \sur{Schumacher}} % 
\author[1]{\fnm{Kilian} \sur{Schwarz}} % 0000-0002-0800-2743
\author[20]{\fnm{Achim} \sur{Streit}} % 
\author[20]{\fnm{R. Florian } \sur{von Cube}} % 0000-0002-6237-5209
\author[4]{\fnm{Rodney} \sur{Walker}} % 0000-0001-8535-4809
\author[9]{\fnm{Cyrus} \sur{Walther}} % 0000-0001-7936-0057
\author[17]{\fnm{Sebastian} \sur{Wozniewski}} % 
\author[21]{\fnm{Kai} \sur{Zhou}} % 

\newcommand{\affilfontsize}{\small}

\affil[1]{\affilfontsize\orgname{Deutsches Elektronen-Synchrotron DESY}, \orgaddress{\city{Hamburg}, \country{Germany}}}
\affil[2]{\affilfontsize\orgname{Max Planck Institute for Radio Astronomy}, \orgaddress{\city{Bonn}, \country{Germany}}}
\affil[3]{\affilfontsize\orgname{Universit\"at Heidelberg}, \orgaddress{\city{Heidelberg}, \country{Germany}}}
\affil[4]{\affilfontsize\orgname{Ludwig-Maximilians-Universit\"at M\"unchen}, \orgaddress{\city{M\"unchen}, \country{Germany}}}
\affil[5]{\affilfontsize\orgname{Justus Liebig University Gie\ss en}, \orgaddress{\city{Gie\ss en}, \country{Germany}}}
\affil*[6]{\affilfontsize\orgname{RWTH Aachen University}, \orgaddress{\city{Aachen}, \country{Germany}}}
\affil[7]{\affilfontsize\orgname{Max Planck Institute for Astrophysics}, \orgaddress{\city{Garching}, \country{Germany}}}
\affil[8]{\affilfontsize\orgname{Forschungszentrum J\"ulich GmbH}, \orgaddress{\city{Garching}, \country{Germany}}}
\affil[9]{\affilfontsize\orgname{TU Dortmund University}, \orgaddress{\city{Dortmund}, \country{Germany}}}
\affil[10]{\affilfontsize\orgname{Friedrich-Alexander-Universit\"at Erlangen-N\"urnberg}, \orgaddress{\city{Erlangen}, \country{Germany}}}
\affil[11]{\affilfontsize\orgname{University of Virginia}, \orgaddress{\city{Charlottesville}, \country{Virginia, USA}}}
\affil[12]{\affilfontsize\orgname{University of Tokyo}, \orgaddress{\city{Kashiwa}, \country{Japan}}}
\affil[13]{\affilfontsize\orgname{Bergische Universit\"at Wuppertal}, \orgaddress{\city{Wuppertal}, \country{Germany}}}
\affil[14]{\affilfontsize\orgname{Heidelberg Institute for Theoretical Studies}, \orgaddress{\city{Heidelberg}, \country{Germany}}}
\affil[15]{\affilfontsize\orgname{University of Bonn}, \orgaddress{\city{Bonn}, \country{Germany}}}
\affil[16]{\affilfontsize\orgname{Universit\"at der Bundeswehr M\"unchen}, \orgaddress{\city{Neubiberg}, \country{Germany}}}
\affil[17]{\affilfontsize\orgname{Georg-August-Universit\"at G\"ottingen}, \orgaddress{\city{G\"ottingen}, \country{Germany}}}
\affil[18]{\affilfontsize\orgname{University of T\"ubingen}, \orgaddress{\city{T\"ubingen}, \country{Germany}}}
\affil[19]{\affilfontsize\orgname{Albert-Ludwigs-Universit\"at Freiburg}, \orgaddress{\city{Freiburg}, \country{Germany}}}
\affil[20]{\affilfontsize\orgname{Karlsruhe Institute of Technology}, \orgaddress{\city{Karlsruhe}, \country{Germany}}}
\affil[21]{\affilfontsize\orgname{Frankfurt Institute for Advanced Studies}, \orgaddress{\city{Frankfurt}, \country{Germany}}}
\vspace*{1cm}

\abstract{Given the urgency to reduce fossil fuel energy production to make climate tipping points less likely, we call for resource-aware knowledge gain in the research areas on Universe and Matter with emphasis on the digital transformation. A portfolio of measures is described in detail and then summarized according to the timescales required for their implementation. The measures will both contribute to sustainable research and accelerate scientific progress through increased awareness of resource usage. This work is based on a three-days workshop on sustainability in digital transformation held in May 2023.}




\maketitle

\section{The challenge}
\label{sec:intro}

\textit{Climate change is real}.
What numbers tell us since many years becomes more and more tangible for everybody in events such as extreme weather conditions, floods, or wild fires that come closer and happen more frequently. This makes it harder and harder to ignore the consequences that our actions as human species have on the conditions for life on our planet.

The discussion often focuses on an increase of average temperature by $1.5$ or $2$ degrees. But this number does not convey the severeness the climate change will have. Each human induced change of the climate increases \textit{the probability of reaching a tipping point} that leads to an irreversible and uncontrolled development. A so far underestimated risk due to global-warming-induced weather extremes like droughts, heat waves and torrential rains may be the reduction of global food production and the resulting expansion of regions with hunger, migration, and conflicts~\cite{IPCC_2023_Synthesis,IPCC_2014_WGII,Kornhuber23}.

We can decide to either accept those risks or mitigate them. If we choose the latter this means we have to \textit{drastically reduce greenhouse gas emissions}. Here it is important to note that the \textit{integrated emission} matters, such that the differential emission per year needs to disappear. Therefore time is a critical factor.

\begin{figure*}[ht]
\begin{center}
    \includegraphics[width=0.95\textwidth]{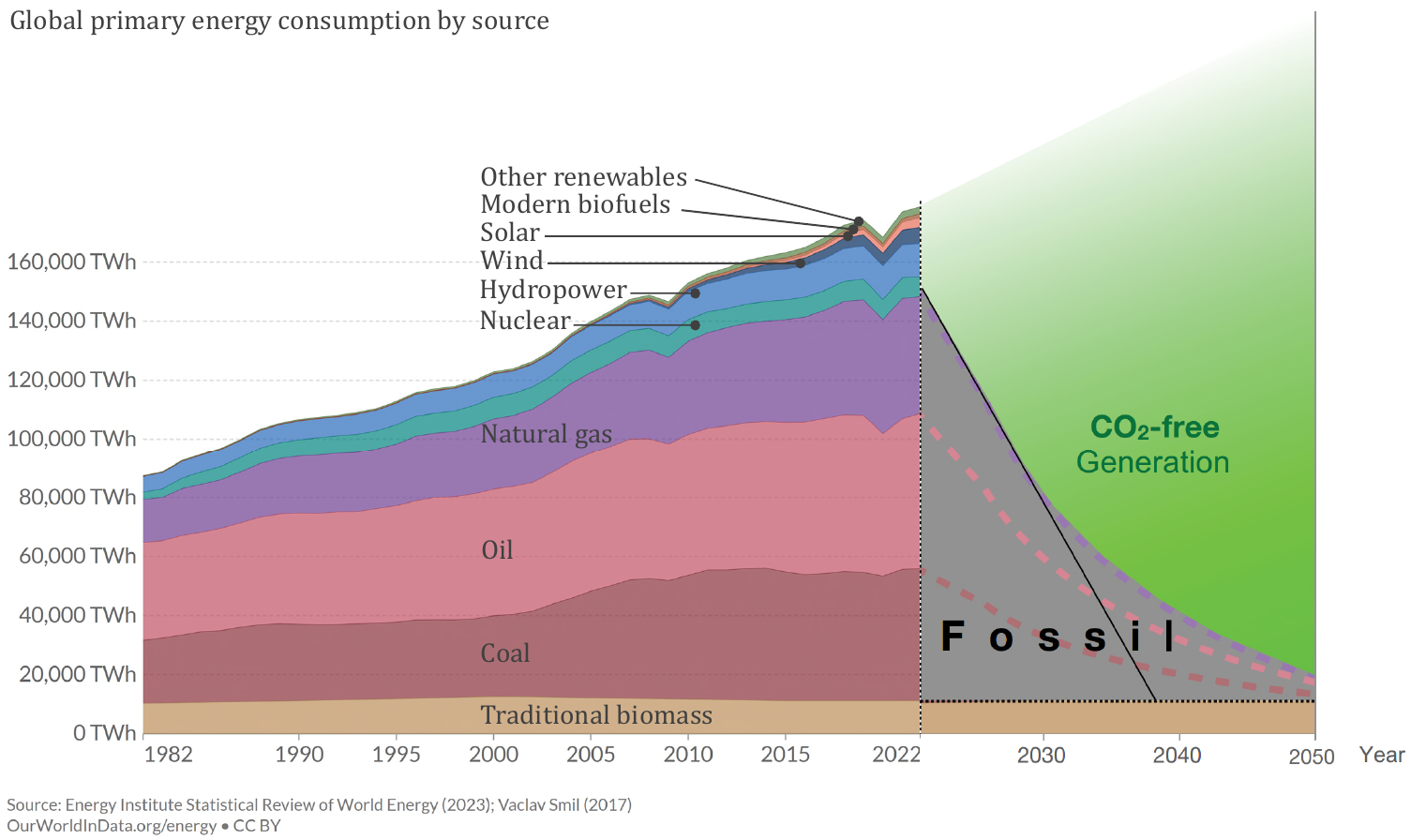}
    \caption{\label{fig:energy}Graph based on the evolution of the worldwide energy production by sources \cite{owidenergy} and projection with required savings according to the Paris agreement by \cite{physikkonkret66}}
\end{center}
\end{figure*}

To meet the goal of the Paris agreement \cite{Paris2015} and reduce the risk of reaching tipping points, the greenhouse gas emissions must be reduced by 50\% within 7 years as illustrated for the worldwide energy production in Fig.~\ref{fig:energy}. Here, the consumption share of data centers is estimated to be about $1\%$ \cite{iea2022}.
The curve labelled \textit{Fossil} is very relevant for ErUM-Data~\footnote{ErUM: Research on Universe and Matter}, the digitization of the research on Universe and Matter in Germany~\cite{ErUM2020,ErUM-Data-Aktionsplan}, because a large fraction of the CO$_2$e~\footnote{Several greenhouse gases CO$_2$, CH$_4$, N$_2$O, HFC, PFC, SF$_6$, NF$_3$ contribute to climate change. Their different effects are normalized to the effect of CO$_2$, the total effect is reported as CO$_2$-equivalent `CO$_2$e' \cite{GWD}.} \textit{footprint of computing in ErUM sciences} comes from the energy consumption for processing and storing data, which is continously growing in many research fields.

If we accept the responsibility arising from our actions and take the Paris agreement seriously, \textit{we as a community must develop a plan} to follow the curve in Fig.~\ref{fig:energy} for the energy consumption of our research. The general transition from fossil to regenerative energies will help and on a time scale of a decade or two there may be an abundant supply of renewable and cheap energy. However, additional savings will be required on short and medium term which can be reached by technological innovations and changes of attitude and behavior.

With these transformations we can on the one hand \textit{contribute to research for sustainability}. On the other hand we must advance \textit{our own research methods to make them sustainable}.

\begin{figure*}[ht]
\begin{center}
    \includegraphics[width=0.8\textwidth]{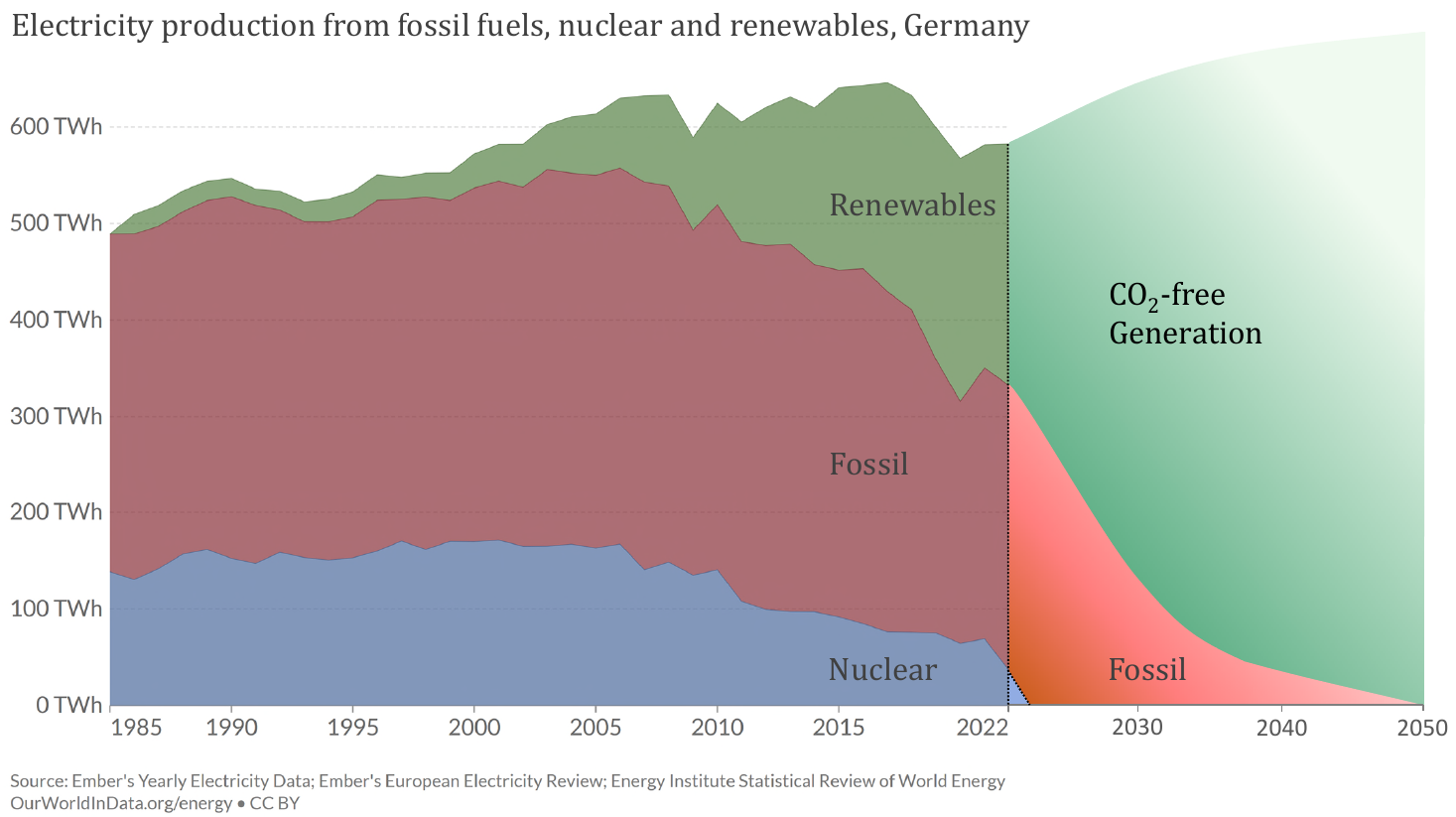}
    \caption{\label{fig:electricity}Shares of the yearly electricity production in Germany \cite{owidenergy} together with a possible extension scenario approximately following expectations in \cite{BMWK,mpg-wind,fraunhofer}}
\end{center}
\end{figure*}

The German government aims at achieving greenhouse gas neutrality by 2045 \cite{BMWK}. While this is an ambitious goal, it still poses challenges in terms of remaining emissions until that year. Figure~\ref{fig:electricity} shows the time evolution of the energy mix focusing on electricity production in Germany, as well as a possible future development based on the targeted expansion of renewables \cite{BMWK,mpg-wind,fraunhofer} which takes into account both the end of nuclear energy in 2023 and the phase-out of coal by 2038.

%ME notes: wind generators deliver 2300h per year compared to 8700h per year impying 25% efficiency (\cite{mpg-wind} or https://www.mpg.de/12699552/windenergie-strom-ertrag), such that adding 23 GW between 2022-2025 will lead to 50 TWh in 2025. Adding 2026-2030 another 66 GW implies 152 TWh.
%ME notes: solar power \cite{fraunhofer}. In 2021 57 GW had been installed that devivered 48 TWh. During the year with 8700h, instead of 57GW*8700h= 496 TWh, about 48 TWh were delivered equivalent to 10%. Adding 45GW between 2022-2025 will lead to 40 TWh in 2025. Adding 2026-2030 another 95 GW gives 83 TWh.
%ME notes: combining wind and solar extensions we expect that from 2022 until 2030 we add 325 TWh compared to the situation in 2022 where we had about 220 TWh from renewables. Thus, thus for 2030 we estimate 545 TWh from renewables out of 658 TWh expected \cite{fraunhofer} consistent with  approx. 80% as expected from the government.

The challenge of sustaining both, our own research and the conditions for life on our planet, is huge, and particularly difficult for the ErUM communities. An extensive compilation of generally relevant aspects for ErUM-related sciences can be found in reference~\cite{Banerjee2023}. In this work we will focus on aspects of digital transformation. Remarkable advances in compute power, data storage, and algorithm development have catapulted the capabilities of modeling, analysis, and simulations to unprecedented levels. In addition, the `fourth paradigm' \cite{hey2009fourth} with its data-centric methodology combined with new artificial intelligence (AI) methods enables researchers to tackle and decipher major scientific challenges. On top of that, there is the increasing sensitivity of experimental instruments and a corresponding increase in data rates and volumes. In light of all this, there are still ideas on how to address the seemingly contradictory overall challenge of preserving both nature and research-driven knowledge gain.

In this document \textit{we present practical measures} in the areas of data and software management, algorithms and artificial intelligence, as well as computing infrastructure discussed at a workshop of ErUM-Data scientists \cite{sustainability_workshop2023}. Twelve questions were formulated in advance which gave a structure to the work and discussions (see Appendix~\ref{app:questions}). The measures can be the basis for the development of a realistic plan for the required reduction of greenhouse gas emissions in the next years while having sufficient computing resources for our research. There are three main approaches.

First, \textit{energy can be saved} by avoiding or reducing unnecessary computations, e.g. by reusing existing results. This requires in particular FAIR~\footnote{FAIR: findable, accessible, interoperable, reusable.} data \cite{FAIRdata} and FAIR research software \cite{FAIRresearchsoftware} as well as an increased \textit{awareness to properly balancing knowledge gain with resource usage}.

Second, we can \textit{increase the efficiency of the calculations} that are required for our research without tapping into the classical rebound trap of using efficiency gains to do more calculations. The measures range from classical or automated code optimizations to new technologies such as new computing architectures and artificial intelligence.

Third, we can \textit{reduce the} CO$_2$e \textit{emissions} caused by our computing. This is mainly in the hand of the computing providers and includes measures such as locating sites close to renewable energy suppliers and reusing the produced heat.

To be able to implement the practical measures discussed in this document certain \textit{competences} need to be acquired and the \textit{right support and incentives} need to be provided by funding agencies, research centers, and universities. The acceptance and execution of the technical and non-technical measures in everyday scientific research will demand a difficult, but unavoidable transition period. \textit{Awareness at all levels is a critical prerequisite for a successful transition}.

The portfolio of measures concerns six areas which are detailed in the following sections: smart data transformation, practices in software development and data analysis, algorithms including artificial intelligence, computing with renewable energies, education and training in sustainable research, as well as funding and institutional support.

\section{Smart data transformation}

Data transformation constitutes an essential aspect of empirical sciences, and advancements in technology are enhancing both its capabilities and complexities, alongside resource utilization. We refer to transformed data designed and susceptible to further exploitation as smart data. As experiments achieve higher resolutions, larger sample sizes or sample quantities, the volume of raw data necessitating processing and analysis increases. 

An example are high-energy particle collider experiments, where permanent storage of all raw data is not feasible since several decades. Automated algorithms (`trigger') are designed to identify the raw data with most scientific value and only retain this data for storage. Nevertheless, current technical and economical boundaries are exploited to maximize the scientific revenue. For example in case of experiments at the Large Hadron Collider at CERN in Geneva, this results in the order of $100\,\mathrm{PB}$ of data being stored per year.

In the field of astrophysics all data taken by telescopes were archived traditionally and evaluated repeatedly. As computing and algorithms evolved, sensitivity to previously unnoticed features improved such that breakthroughs have appeared from reprocessing of archival data (e.g. \cite{Lorimer:2007qn}). Such research with archived data may be particularly resource-efficient. In the future, storing all data will no longer be possible as newer and more sensitive state-of-the-art facilities such as the Square Kilometre Array Observatory will produce data rates in the range of Terabits per second that can no longer be stored, but must instead be analyzed immediately and deleted after processing.

In photon and neutron sciences, fast automated sample exchange and qualitatively improved detectors cause similar effects. As pioneering examples, `stream based models' are explored by not storing raw data at all, and performing all processing online \cite{SRI-conference-2023}.

Dealing with such large amounts of data naturally requires significant compute resources with non-negligible footprints in energy consumption (see Section~\ref{sec:hardware}). Hence, smart data transformation becomes increasingly important, especially in view of further envisioned detector upgrades and new science facilities. 

\textit{For achieving our sustainability goals}, teams with in-depth knowledge of the relevant experiments are needed to develop comprehensive data management strategies. Their efforts are in direct relation to the goals of the NFDI consortia \cite{punchfornfdi, daphnefornfdi, fairmat}. The teams need to provide a data life cycle with different phases of data use, from instant analyses after data recording to completely new analyses at a much later time (long-term archiving). In the early stages, storage capacities could be substantial to maintain some level of data redundancy for backup and convenient access. In later phases, the volume of data retained could be gradually reduced over specified time intervals.

For the content assessment of the data, far-reaching expertise is needed to define the exact objectives of the data storage. The spectrum ranges from timely thorough data analyses, to reproduction of analyses, later cross-verification or subsequent use in completely new contexts. For the latter, potential future use cases should be anticipated as best as possible. 

It is thus mandatory to define which data should be stored in which formats. The criteria under the overarching FAIR data principle  \cite{FAIRdata} for data preservation should also be followed in Universe and Matter research. 

For data to be reused, their meaning must be self-explanatory to scientists and convey information about the conditions under which the measurement data were obtained. Such metadata on location, time, device, operating conditions, instrument calibrations, etc. play a critical role in understanding and processing the data in subsequent phases.

Another aspect of discoverability concerns the descriptive language of the data. Within individual research communities, there are often specific terminologies and vocabularies whose linguistic nuances evolve over time and thus can present significant obstacles to the use of archived data. In some research areas, ontologies have been developed to describe preserved data. This approach creates a standardized lexicon and thus simplifies the matching between the description provided with the data and a search query.

Finally, workflows are important for smart data transformation, referring to software and algorithms as well as processes involving humans. In several areas of the ErUM community, it has become common practice to take snapshots of processed data at intermediate stages in workflows throughout the data processing chain and store them temporarily. This approach allows for later extensions and repetition of subsequent steps without having to reinitiate the entire data processing sequence. This efficient strategy saves time and computational resources, albeit at the cost of temporarily increased storage requirements.

With the goal of achieving scientific results as soon as possible, there is an immediate motivation for the efficient choice of such snapshots, and they should improve energy efficiency. Orchestrations of workflows are also being tried in large collaborations where the same data on different aspects are analyzed by individual research teams. However, it requires a strong commitment and dedication to collaborative efforts to prioritize resource efficiency through joint action, especially in highly competitive contexts.

\vspace*{3mm}
\noindent
As a summary, we note these actions for the topic of smart data transformation:
\begin{itemize}
\item Make data FAIR to promote reuse, which can be particularly resource efficient.
\item Reduce and compress data, having the anticipated scientific value of the retained information and the resource requirements in mind.
\item Optimize the choice of storing intermediate results against re-calculating them.
\item Optimize job orchestration and scheduling in workflows.
\end{itemize}

\section{Software engineering and data analysis \label{sec:software}}

The relevance of human resources invested in software development is increasing in view of the ever-growing volumes of data and the corresponding computing power required. Once written, code is applied to increasingly more data. Therefore, prioritizing software optimization is more than justified.

A common practice in dealing with research data is to code an exploratory analysis that investigates the potential of the data with respect to a scientific question. In case of success, a re-engineering of the used software into a professional structure is mandatory to consolidate and enable possibilities of enhancing the code base. Furthermore, especially in the case of complex data analyses with many intermediate steps, it is essential to introduce a workflow management system to ensure both the reproducibility of the analysis and to avoid unnecessary recalculations \cite{Rieger:2020ytt}. Often, this revising step is skipped and it remains with what is popularly known as `spaghetti code' together with the researcher who has memorized the order of the calculations for himself, inevitably leading to `abandonware'. Experience shows, however, that a solid structure of the research software and the use of workflow managers accelerates iterative review processes and thus the publication of the results.

Accordingly, adherence to good software development practices has the potential to save human and energy resources through excellent code quality. The FAIR research software seal \cite{FAIRresearchsoftware} now exists in this area as well. Community-maintained, easily accessible open source code is generally more efficient than local stand-alone implementations. Effective coding practices of jointly developed software include tracking systems, version control, rigorous testing, and benchmarking (see, e.g. energy per luminosity in GWh/fb$^{-1}$ which is particularly suited for particle physics~\cite{metrics}). Furthermore, modularization and reuse of code, possibilities for parallelization of computations (vectorization), comprehensive documentation, and continuous integration are important.

Generally, energy efficiency should play a central role in software development, along with broadly-accepted software-quality metrics. First of all, unnecessary computations should be avoided. This is helped not only by workflow managers, but users can also execute initially small-scale validations and monitor execution for early detection of software or parameter problems. Code developers can implement sanity checks on configuration problems and input data. In this way, wasting energy consumption by large, unsuccessful runs can be avoided.

Great help in conserving resources comes from using established runtime optimization techniques through automated parallelization, adaptation to new CPU architectures (i.e. ARM, RISC-V, etc.) and Co-Processor architectures, vectorization, memory layout optimization, and the use of special compiler flags. All such measures generally lead to more energy-efficient code and are also being further developed in expert groups in the Universe and Matter research area \cite{MatthesWZWHB17}.

In order to intensify the necessary awareness for resource consumption, tools for monitoring of resource consumption, trainings on the conscious use of resources (see Sec.~\ref{sec:culture}), practical optimizations of code as well as their computational processes need to be developed and made available. 

Ultimately, given the pivotal role of high-quality scientific software in the research landscape, its development and maintenance should receive increased and appropriate valuation. This could encompass recognition through publications and citations, funding allocation, enhanced access to infrastructure resources, and the establishment of dedicated career paths.

\vspace*{3mm}
\noindent
Summarizing practices of software engineering, we note these actions:
\begin{itemize}
\item Make software FAIR and reliable by following good software development practices and ensuring sustainable support.
\item Design software for optimized energy consumption.
\item Use workflow management to make processing FAIR.
\item Continue research on potential of new technologies for efficient use of resources.
\end{itemize}

\section{Algorithms and artificial intelligence \label{sec:algoAI}}

The energy efficiency of algorithms is crucial for the reduction of computing resources and obviously closely linked to adherence to the above-mentioned good software development practices. Often, there are already optimized algorithms in open-source standard libraries or domain-specific libraries that can be utilized. A recent example is experiments with Big Data in the form of millions of events, in which the formerly sequential event loop has been replaced by parallel processing of large event chunks \cite{dask,coffea,Eich:2022ctj}. Beyond this, new challenges may soon arise from dynamic sustainable energy supply of computing centers, to which algorithms may also have to be optimally adjusted.

The rapid progress in the field of artificial intelligence (AI) is opening up completely new possibilities. Machine learning is already accelerating ErUM research in theoretical predictions, in simulations, and in the analysis of experimental data. Moreover, innovative data analyses are becoming possible via these data-driven modeling techniques, in which the physics potential of an experiment is exploited to a much greater extent than originally expected (e.g., \cite{PierreAuger:2023gmj, IceCube:2023lxq}).

From many more examples, we mention here also closed-loop experiments in X-ray and neutron sciences with their direct feedback after data taking from quasi-instantaneous data analysis and control of the experimental parameters in film growth \cite{Pithan:ju5054}, and more efficient searches in a dedicated phase space for dispersion relations of phonons or magnons \cite{Astrid2023}. Finally, we expect that in the near future more and more tasks will be performed by scientists using AI tools.

We argue that the full potential of machine learning algorithms should be prioritized, respecting the goal of reducing energy consumption in ErUM research. The focus should therefore be on developing and deploying models to perform tasks that have the potential of great benefit to scientific progress, or are particularly expensive in terms of computational resources. Given the limited time window for reducing the CO$_2$e footprint of ErUM computing, developments in areas with high potential for reducing energy consumption must be advanced early. 

However, since training machine learning models also leaves a CO$_2$e footprint, the resources consumed in training should be evaluated and documented for transparency. This should include all training performed, including those to optimize the models. Energy consumption during the inference period is likely to be low in comparison, but should also be documented and compared to the consumption of established alternative solutions.

In light of recent breakthroughs in generative models, this is one area where the deployment of machine-learning solutions has large potential for fields of ErUM that rely on large samples of simulated data. One such example is the detector simulation at the LHC experiments, which at the ATLAS experiment for example consumed 38\% of the total CPU resources in 2018~\cite{Calafiura:2729668} and for which a deep generative model is being deployed that is $\mathcal{O}(500)$ faster than the simulation with {\sc Geant4}~\cite{ATLAS:2021pzo}.

Another promising area is the adaptation of large pre-trained models that are only refined for the specific ErUM applications. Thus, refining such pre-trained models would not only require less computational resources in training, but also enable smaller ErUM training datasets. In general, the reusability of previously trained machine learning models has the potential for more efficient training in ErUM research.

Pre-trained models include recent AI developments in terms of large language models, which are available to the general public as new and powerful AI tools. These tools have great potential to increase work efficiency in ErUM research, for example, by helping with documentation tasks. This includes computer code documentation, which is time-consuming and therefore often neglected.
Good code documentation has the potential to increase the reusability of code in general and of efficient code in particular. However, large language models can also be used to directly suggest more energy-efficient algorithms. These new capabilities are directly related to the identified needs in Section~\ref{sec:software}.

It can hardly be overemphasized that scientists have a key role in the choice of algorithms. Their decisions on the use of self-coded algorithms, library algorithms, or artificial intelligence algorithms are crucial to the runtime of the jobs and their resource consumption. Considerable expertise is required especially for successful deployment of new AI algorithms, i.e., deployment that is energy efficient and leads to scientific progress. 

Therefore, the strategic use of AI tools in the scientific workflow needs prioritization between the use of human and computer resources and in the expected value of knowledge gain. As a consequence, it needs to be decided where AI should be used sensibly and where not. For example, for some problems, comparable performance could be achieved by human reasoning instead of resource-intensive machine learning training. When AI-generated code or machine learning is used for ErUM research, it always requires extremely careful validation by scientists. Thus, it is clear that human interaction will remain a pillar of scientific progress.

\vspace*{3mm}
\noindent
We see the following measures to be central in the area of algorithms and artificial intelligence:
\begin{itemize}
\item Continue research on potential of AI or other new technologies.
\item Include particularly promising applications of generative and pre-trained models.
\item Expand detailed monitoring and documentation of energy consumption and CO$_2$e footprint in training and inference.
\item Use already optimized algorithms in open-source standard libraries or domain-specific libraries.
\end{itemize}

\section{Computing and infrastructures \label{sec:hardware}}

In this section, we focus on sustainability related to computing hardware and its operation. Discussions of practical measures involve adjusting computing in space and time to the availability of renewable energy, reusing the generated heat, and extending the operating lifetime of hardware. Implementing such measures requires a comprehensive information flow between the stakeholders involved.

Greenhouse gas emissions are widely classified in so-called scopes (e.g. \cite{nationalgrid}). There are no directly produced emissions from computing for research on the Universe and Matter (scope 1). However, indirect emissions are produced by operating the data centers with electricity that is not produced from renewable sources (scope 2). Finally, indirect emissions arise from the entire value chain, starting with the production of buildings and computer systems and later with their disposal (scope 3).

\paragraph{Renewable energy (scope 2):}
First, with regard to scope 2, an essential component of sustainable data processing is a detailed overview of the power consumption of the various systems and services, as well as a detailed accounting of past or planned activities. Data centers generally have detailed measurements and records of the power consumption of their various systems, although this information is generally only available upon individual request and is not yet directly accessible online in a comparable format. 

For the annual electricity consumption of large German data centers, we refer to the compilation results in Tab.~\ref{tab:RZenergy}. For comparison, $1$ GWh roughly equals the electricity demand of $1,000$ single-households per year in Germany.
\begin{table}[htb]
    \centering
    \begin{tabular}{lll}
    \toprule
    Computing center & Electricity/GWh & Ref. \\
    \midrule
    MPCDF Garching (2022) & $43$  & \cite{MPCDF-2021}\\
    LRZ Garching  (2021)  & $33$  & \cite{LRZ-2021}\\
    HLRS Stuttgart (2021) & $32$  & \cite{HLRS-2021}\\
    JSC Julich (2012/13) &  $34$  & \cite{JSC-2012}\\
    \bottomrule
    \end{tabular}
    \caption{Annual electricity consumption of large computing centers}
    \label{tab:RZenergy}
\end{table}

An example of the approximate breakdown of energy consumption among the major components of the CERN data center is $55\%$ data processing, $21\%$ disk storage, $2\%$ tape storage, $5\%$ network, and $17\%$ services \cite{Campana:chep23-wlcg-energy}. Details of these numbers can be found in the Appendix~\ref{app:breakdown}. During the $13$-year operation of the Large Hadron Collider, power consumption has been fairly constant, although computing capacities have increased by a factor of $6$ from 2012 to 2023. The high service share comes about because of the special role of the CERN computing center in the interconnection of about $170$ computing centers in the Worldwide LHC Computing Grid `WLCG' \cite{Campana:chep23-wlcg-energy}, where CERN operates a large part of the central services. 

A next big step toward sustainable computing would be to place data centers near sustainable energy sources. Renewable energy supplies generally consist of solar panels and wind farms, but also biogas power plants.

Practical approaches with data centers directly at the producers already exist in Texas in the USA, whose renewable energy capacity will reach approx. $70$ GW by the end of 2023 \cite{grimshaw}. Transporting electricity is a major problem due to lengthy permitting processes for transmission lines. So the idea of moving electricity consumers to the point of generation was born. While this is difficult for most industries, scientific computing is an ideal candidate for this relocation. On the hardware side, data centers only need power and fiber for the network. The user side of scientists typically computes in batch mode, usually considering temporary interruptions due to insufficient sun or wind and thus moderate delays in the computations acceptable.

For Germany, a comparable data center scenario could be built near the North Sea, where most of the wind is available and most offshore wind farms have either been built or are in the planning stages. Substantial computing infrastructure could be built as green-field sites, as is already explored commercially \cite{windcloud}. It is advisable to keep sufficient storage data capacity along with the computing resources to reduce network latency issues when data throughput is large.

Since in many communities experiments are carried out worldwide and as international collaborations, data of interest to scientists working in Germany could be relocated to such data centers in the north. The cost and effort of laying the required fiber optic cables for the networks likely do not dominate. Transformed data with lower volumes could optionally go to the universities for further processing.

Overall, it makes sense to establish a joint ErUM science cloud initiative timely, starting with moderate equipment and scaling up once significant funding has been successfully acquired. Even in the initial pioneering phase, there are many aspects to develop and explore, as we will discuss below.

When setting up a data center near the power producers, the following criteria should be taken into account. Since the energy comes from renewable sources, the performance of a processor per invested units of electricity (flops per watt) plays a subordinate role. Thus, initially hardware could be operated that does not belong to the latest generation and can therefore be obtained at relatively low cost. 
%As a side effect, the footprint of newly produced hardware is reduced (scope 3). 

The key challenge for data centers is the dynamics in supply from renewable energies  \cite{ZCCloud,grimshaw,DESY-RF20}. Data centers must be able to dynamically ramp up or down computing resources or shift workflows as needed. Importantly, centers must be equipped with valid forecasts of weather conditions and expected energy supply (see below paragraph on information flow and middleware).

There are two relatively straightforward ways to reduce power consumption in a data center: First, one can reduce the CPU clock rate. In principle, this can be done immediately; jobs are only slowed down, but continue to run. Studies show that power consumption of compute nodes can be reduced by $50\%$, with a corresponding slowdown in processing \cite{Walker:sust-ws}. 

Second, one can hibernate or power down entire nodes. The order of nodes could follow a prioritization list based on the operating age of the hardware and thus their efficiency. Shutting down CPUs works within seconds, while restarting can take several minutes depending on the memory requirements and the speed of the IO system \cite{grimshaw}.

However, it must be ensured that running jobs are stopped properly without losing the results achieved up to that point. Ideally, this can be achieved by so-called checkpointing, i.e. the entire program state is stored on disk and can be resumed later. In practice, this feature is challenging for data processing jobs with many open connections to external services. If the processing consists of repetitive, independent steps, as is the case with event processing in particle physics, an alternative is event-level check-pointing, i.e., after each processing step, the output is stored in its entirety. In case of an interruption, processing can be resumed after the last processed event, so that only little CPU time is lost.

However, a minimum energy supply to the data center must be guaranteed at all times. The servers and network switches of the data center should run continuously. A data center with CPU and storage as in use for WLCG requires about 25\% of power for hard-disk drive storage (HDD). To avoid damage, these servers should not be shut down frequently, resulting in a continuous power requirement. Therefore, energy storage options should also be planned for the data center from the beginning. Depending on the environment, these could be accomplished for example by batteries, water storage, energy to gas plants, flywheels, and bidirectionally charging electric vehicles.

Efforts to optimally use the supplied energy for computing centers are rated by Power Usage Effectiveness (PUE) which describes the total amount of energy used by a center compared to the energy delivered to the computing equipment. For research infrastructures such as CERN in Geneva, the Prevessin Site reaches PUE=$1.1$ \cite{Walker:sust-ws}, the National Renewable Energy Laboratory in Boulder (USA) reported annualized PUE=$1.036$ \cite{NREL2023}, and the German HPC center LRZ measured a PUE of $1.06$ for the SuperMUC-NG system \cite{LRZ-2022}.
Beyond energy-supply discussions, modern computer chips have very high heat output per unit area, exceeding that of a conventional induction stovetop \cite{struckmeier}. In terms of sustainability, it is imperative to use the dissipated heat, for example, to provide heating and hot water for the nearby buildings. Also, residential units are provided with data centers where best PUE values were achieved for new residential buildings (PUE=$1.024$), but also far-reaching improvements were obtained in conversions of e.g. high-rise buildings from PUE $\sim2$ to PUE=$1.27$ (both in \cite{struckmeier}).

\paragraph{Hardware lifetime (scope 3):}
A subject of its own are the sustainability issues for the above-mentioned scope 3, which takes into account the CO$_2$e footprints during manufacturing and disposal. It is not easy to get exact data for individual hardware components. There are studies that put the manufacturing CO$_2$e footprint at $20-30\%$ of the total CO$_2$e footprint \cite{dell:hw,gupta2020chasing}. Obtaining corresponding numbers for e-waste turns out to be challenging.

At this scale, hardware lifetime is a relevant issue. Complementary variables are power consumption and lifetime. Data center equipment runs $24\times 7$ continuously for approximately $5$ years, containing the aforementioned approximately 25\% of fixed carbon. Extended life in certain mission areas - i.e., larger science centers with dedicated and experienced staff could explore more versatile life extension options that result in $7$--$10$ - or even more - years of operation which is worth when using renewable energies.

However, there is also a finger pointing at typical personal devices such as desktops, laptops, smartphones, etc. Because of the much shorter, integral operating times and sophisticated energy efficiency, the CO$_2$e footprint of manufacturing here goes in at about 75\% bound carbon. Accordingly, life extension for devices for individual use needs to be seriously considered.

Recent announcements indicate that industry is now putting a stronger focus on energy efficiency in data center devices as well \cite{sert2}. The corresponding component-based replacement and operation of data center devices with renewable energy are clearly going in the right direction to address the challenges of sustainability scopes 2 and 3.

\paragraph{Information flow and middleware:}
Successful sustainable computing in the area of research on Universe and Matter requires comprehensive information flows between energy providers, data centers and users. Furthermore, conceptually new middleware tools are needed for dynamic operation of the data centers. These issues are addressed here.

First, monitoring of resource consumption is central to further advancing energy efficiency. What is the consumption per job, per user, per publication, or the entire data center? Details such as the resource profile within a computing job are also required. With such tools, cost-benefit analysis and optimization efforts are made possible in the first place. 

Today, only a few researchers, groups and departments receive information on the resource usage of their computations. One example is the National Analysis Facility (NAF) at DESY, which recently set up a monitoring service to determine the energy consumption of a job in the central processing system and record this information in the log file. The service also provides a CO$_2$e estimate based on the current energy mix at the data center site. Such systems need to be transferred to provide every scientist with comprehensive energy and CO$_2$e reporting.

The development of easy-to-use tools for explicit measurement and profiling of energy usage and CO$_2$e footprint for developers and end users is non-trivial. First, the CO$_2$e footprint depends on the energy resource used, which is known to the energy provider only. The relation between energy consumption and program runtime is not necessarily strictly linear, since modern architectures have dynamic power and frequency scaling. In addition, a poorly optimized GPU implementation may run faster but consume more energy than a CPU version due to differences in thermal design performance.

A bidirectional flow of information between users and the data center is also necessary. Users should be able to estimate the requirements of their job types, at least approximately. Detailed accounting information about the energy consumption and CO$_2$e of each data processing job, production task or data transfer would be important and useful information. Similarly, users should be able to provide information about what specifications their algorithms tolerate and the dynamics with which their algorithms can respond to power shortage situations. Furthermore, it should be possible to define whether certain delays in batch jobs or even reduced job numbers are compatible with their quality-of-service requirements. An example are High-Throughput Computing applications of WLCG data processing, which are -- within boundaries -- less time-critical. In other words, for a major campaign that lasts many days, delays of a few hours are not critical.

Conversely, users need information from the data center on what the supply situation is -- whether, for example, computing resources will be slowed down or even temporarily shut down -- and, finally, what the forecasts are for job completion. In general, one can expect downtime to be acceptable as long as it is predictable on a daily basis. Users would not submit jobs that cannot be completed in the foreseeable future.

Second, we can expect that in the medium term, consumer energy prices will be time-variable, and will depend on the availability of renewable energies and on the overall energy demand. In order for data center operators to perform energy-dependent work planning, it is important to obtain information from energy providers about current pricing and availability of renewable electrical energy, e.g., whether there is an oversupply or undersupply. In addition, forecasts for supply, including weather information and price stability, are needed. Conversely, the energy provider needs the requirements of the data center for its own work planning in a timely manner.

To enable overall asynchronous communication flows, one possible approach to the required information exchange between all stakeholders could be to set up a centralized and scalable or distributed monitoring infrastructure where data is collected in real time and where everyone can query the required information using standard protocols and APIs as well as standardized communication content.

Beyond the aforementioned information flows, data centers require conceptually new middleware tools that enable efficient energy-aware scheduling ~\cite{Kocot23sched} based on the overall situation of user demand and energy availability and implement dynamic CPU/GPU power modulation and load limitation on existing systems. Good results can be achieved with applications that adapt to changes in resource availability (so-called malleable jobs~\cite{D_Amico_2019}). Also users need new middleware tools to hand in their job requirements and receive reports.

For longer-term periods of low sun or low wind, energy storage technologies must also be developed and deployed. A scenario must be developed on how to react in such situations. Historical weather data could be used to define storage capacities: how often and for how long do such situations occur. 

A possible contingency procedure could be to first use up the energy storage, then freeze jobs or only run high priority jobs for a certain period of time, and finally use non-renewable energy. The guiding principle should be to keep the CO$_2$e footprint of data center operations as minimal as possible \cite{DESY-RF20}.

\vspace*{3mm}
\noindent
From the topic of computing with renewable energies we summarize the following demands:
\begin{itemize}
\item Monitor and report energy consumption at job level including resource profiling within the job.
\item Monitor and report energy consumption at site and project level, provide information of the individual use per scientist/project/publication.
\item Extend monitoring of resources beyond CO$_2$e (water, material etc.).
\item Consider carbon footprint for all planned investments and project plans.
\item Adjust computing in space and time to the availability of renewable energy, e.g. computing centers close to off-shore wind parks with a job scheduling using only or mainly the surplus of renewable energy available at a given time.
\item Develop software and middleware that can respond dynamically to the availability of energy.
\item Optimize power usage effectiveness.
\item Re-use produced heat.
\item Adjust hardware lifetime considering emissions due to procurement and operation.
\end{itemize}

\section{Developing a culture for sustainable science in the ErUM communities \label{sec:culture}}

There is no disputing that remarkable advances in computing power, data storage, and algorithm development have greatly accelerated the capabilities of modeling, analysis, and simulation to unprecedented levels. This transformation has opened new frontiers of scientific understanding and progress that were impossible at the turn of the century. Especially data-centric methodology has revolutionized numerous scientific fields, enabling researchers to tackle and decipher grand challenges by studying systems at multiple levels with unprecedented precision (see Section~\ref{sec:algoAI}). 

However, as a result of this development, there has been a significant increase in energy consumption, which needs to be accounted for in relation to scientific progress, so there is an urgent need for comprehensive, strategically designed education that covers all of these areas. 

Since the early 2000s, academic discussions and surveys \cite{Wiek2011SKC} have already underscored the importance of sustainability-focused education and the vital skills needed to equip the next generation of scientists with sustainability awareness and the aptitude necessary for innovative research and development. In addition, UNESCO describe the World Programme of Action on Education for Sustainable Development (2015-2019, 2020-2030) \cite{UNESCO2020} and offer detailed recommendations and strategies on how to effectively integrate AI into education systems in its comprehensive guide \cite{AI_Education_UNESCO}.

New university courses and workshops designed to empower students and researchers with the knowledge and skills to conduct sustainable digital research can help create a responsible approach to knowledge gain in conjunction with resource usage. Together with the portfolio of measures described in this publication, these educational tracks have the potential to promote informed decision making about the use of computing resources and to support more sustainable practices in scientific research.

Especially when using data-driven methods, a commitment to sustainability is essential. To balance and monitor the seemingly inexhaustible potential of AI, researchers and students must have deep domain-specific knowledge. A solid foundation in mathematics and computation paves the way for the deliberate integration of efficient computation and algorithms, streamlined models, and conscious decision-making throughout the research process. 
This is where it makes sense to inject one’s own intelligence first, rather than reflexively increasing computational power.
A further temptation notes that efficiency improvements trigger a rebound effect that increases the demand for computing resources. Thus, regular reviews and controls may be needed to mitigate this phenomenon.
This and other incentives under development (e.g. ~\cite{Pardis-yHEP}) can certainly be beneficial. Even more robust measures such as CO$_2$e-based fairshares, CO$_2$e budget allocations, and specific reduction commitments (`carrot and stick' strategy) are conceivable.

Thus, for early-career researchers, knowledge transfer, mentorship, resource allocation, and networking are central to cultivating their skills, deep understanding and effective use of AI tools. 
For experienced scientists in leading positions, it is an important task to assess the CO$_2$e footprint of their current research, to develop, implement, and monitor plans to reduce CO$_2$e emissions, and to consider CO$_2$e emissions in future investments and project plans. 
With these approaches and balance between innovation, awareness, and responsibility, we can continue pushing the boundaries of human understanding while preserving the resources we depend on for a sustainable future.

Overall, there are many technical aspects, guidelines, developments, and scientific measures to improve the current scientific work in our research field with respect to the urgency of climate change. However, the success depends on the implementation in the daily work of individual scientists and therefore requires an urgent change in awareness and responsibility of every scientist and science manager.

For the development of educational concepts and to raise awareness, we summarize here key concepts that should serve to refine priorities for performing research sustainably:
\begin{enumerate}
    \item \textit{Balance the knowledge gain of any (computational) work against the resources it needs.}
    \item \textit{Reliability is sustainability, avoiding unnecessary repetitions.}
    \item \textit{Thought-through, well-documented workflows and standardization guarantee reliable results and their re-use.}
\end{enumerate}

A further periodically appearing challenge of working sustainably in ErUM sciences results from project-oriented organization of the funding schemes. The sustainable use of invested resources requires improvements in the long-term organization of knowledge transfer and the exchange of results and methods. For this continuous support, sufficient resources in the areas of software, algorithms, and computation need to be allocated.

\vspace*{3mm}
\noindent
For developing a culture for sustainable science in the ErUM communities we summarize the following measures:
\begin{itemize}
\item Raise awareness of the climate challenge at all levels.
\item Disseminate knowledge of measures to address the challenge.
\item Train scientists in good practices.
\item Strive to become a role model at all levels and help to establish sustainability in everyday life.
\item Enhance awareness of the trade-off between research benefit and climate impact.
\item Perform first, rough energy audit and develop an initial CO$_2$e reduction plan.
\item Regularly review and update the CO$_2$e reduction plan.
\item Consider carbon footprint for all investments and project plans.
\item Include the resources needed for continuous IT support into project planning.
\end{itemize}

\section{Funding and institutional support}

From the perspectives of the federal and state ministries as well as the top management of universities and research centers, there is a great deal of attention being paid to the issue of sustainability. This involves the development of strategic concepts, reporting, measures and funding.

At the level of the German Federal Government, the United Nations \textit{Agenda 2030} \cite{SD2015} has been transferred into a national strategy named \textit{German Sustainable Development Strategy 2021} \cite{DNS2021}. A comprehensive general work on the \textit{Digital Strategy Germany} was published subsequently~\cite{DGS2022,DGS2023} which concerns science and research among many other aspects and is closely linked to the sustainability goals. In more specific contexts, the Federal Ministry of Education and Research (BMBF) developed sustainability measures, e.g. which for science is called \textit{Research for Sustainability (FONA)} \cite{FONA2020}. 

Within the framework programme ErUM \cite{ErUM2020}, the BMBF has strengthened the aspect of sustainability in research on Universe and Matter in their calls for proposals. For example, technological and methodological development work that contributes to climate- and resource-friendly operation of large-scale facilities and experiments can be funded alongside projects (e.g. \cite{BMBF2023}).

In view of all the initiatives and activities to date, however, major and urgent tasks remain for all stakeholders to be mastered together. The many measures described in this work require appropriate support mechanisms and structural alterations, some of which themselves must be sustainable. For example, computing facilities localized at power generators need infrastructure and personnel. Furthermore, an expert group to advise and support scientists for Digital Transformation measures in ErUM could be implemented, e.g. in a new phase of the ErUM-Data-Hub \cite{erumdatahub}. 
%Likewise, careers in the field of sustainable research data and sustainable software and computing need to be established by creating corresponding long-term positions. 
In any case, a sustainable impact can definitely be achieved by creating appropriate long-term positions in research data, software, and computing.

Efforts to lower bureaucratic and legal hurdles for the implementation of sustainable dynamic power generation and transfer are most welcome. They are needed along with new communication channels between power companies, data centers and scientists as consumers to enable practical and pragmatic solutions for the sustainable use and operation of computing infrastructures.

\begin{table*}[ht]
    \centering
    \small{\begin{tabular}{|ll|}
    \toprule
    Item & Call-to-action \tabularnewline
    \midrule
         & Immediately or on \textbf{short time scale} with little effort these measures can be implemented:\tabularnewline
    \midrule
    S1   & Raise awareness of the climate challenge at all levels.\tabularnewline
    S2   & Disseminate knowledge of measures to address the challenge.\tabularnewline
    S3   & Monitor and report energy consumption at job level.\tabularnewline
    S4   & Consider carbon footprint for all investments and project plans.\tabularnewline
    S5   & Enhance awareness of the trade-off between research benefit and climate impact.\tabularnewline
    \midrule
         & On a \textbf{medium time scale} of a few years the following measures can be realized:\tabularnewline
    \midrule
    M1   & Make data FAIR to promote reuse.\tabularnewline
    M2   & Reduce and compress data having the anticipated scientific value of the retained information\tabularnewline      
            & and the resource requirements in mind.\tabularnewline
    M3   & Optimize the choice of storing intermediate results against re-calculating them.\tabularnewline
    M4   & Optimize job orchestration and scheduling in workflows.\tabularnewline
    M5   & Use workflow management to make processing FAIR.\tabularnewline
    M6   & Make software FAIR and reliable by following good software development practices \tabularnewline
            &and ensuring sustainable support.\tabularnewline
    M7   & Design software for optimized energy consumption and provide tools to measure it.\tabularnewline
    M8   & Continue research on potential of AI or other new technologies for efficient use of resources,\tabularnewline
            & but balance gain of research action against resource consumption of these developments.\tabularnewline
    M9   & Monitor and report energy consumption at site and project level, provide information of the \tabularnewline
            &individual use per scientist/project/publication.\tabularnewline
    M10   & Extend monitoring of resources beyond CO$_2$e (water, material etc.).\tabularnewline
    M11  & Train scientists in good practices.\tabularnewline
    M12  & Regularly review and update the CO$_2$e reduction plan.\tabularnewline
    M13  & Strive to become a role model at all levels and help to establish sustainability 
            in everyday life.\tabularnewline
    \midrule
        & A \textbf{longer term} coordinated planning is required for the following measures:\tabularnewline
    \midrule
    L1  & Adjust computing in space and time to the availability of renewable energy,           e.g. computing \tabularnewline
        & centers close to off-shore wind parks with a job scheduling using only or mainly the surplus \tabularnewline
        & of renewable energy available at a given time. \tabularnewline
    L2 & Develop software and middleware that can respond dynamically to the availability of \tabularnewline
        & renewable energy. \tabularnewline
    L3  & Optimize power usage effectiveness.\tabularnewline
    L4  & Re-use of produced heat. \tabularnewline
    L5  & Adjust hardware lifetime considering emissions due to procurement and operation. \tabularnewline
    L6  & Include the resources needed for continuous IT support into project planning. \tabularnewline
    \bottomrule
    \end{tabular}}
    \caption{Call-to-action in digital transformation: portfolio of measures to be taken, ordered in terms of effort and time they take.}
    \label{tab:measures}
\end{table*}

\section{Conclusions}
\label{sec:conclusions}
Based on a three-day workshop on sustainability in the digital transformation, interested colleagues from the ErUM-Data community have developed a portfolio of measures with which our research area will massively reduce the emission of climate gases in the sense of the Paris climate agreement of 2015 \cite{Paris2015} and turn to the use of renewable energies. In doing so, we are planning for a transitional period until electricity will once again be available in abundance through the establishment of a sufficient number of renewable energy sources. 

In light of the dramatic increase in data rates from new or upgraded instruments and the enormous potential for new knowledge gains from developments in data-driven, AI-assisted methodologies, the awareness of each of our scientists is critical during this transition period. Already by applying the portfolio of targeted sustainability measures described in this publication and consciously balancing prospective knowledge gain and resource usage, we expect to achieve efficiency gains that will ultimately accelerate our fields of research. 

In the portfolio, we see six areas: 1) considering sustainability as an important factor in the planning of measurement data transformation, 2) good practices in software development and data analysis, 3) use and development of renowned and efficient algorithms including AI, 4) data center locations at sustainable energy providers along with lifetime extension of hardware, 5) education and training in a responsible approach of balancing knowledge gain and resource usage, and 6) targeted efforts of all stakeholders including funding and institutional support. 

Our ErUM community can respond now and develop a realistic plan for the required reduction of CO$_2$e emissions. We have the tools at hand or can develop them. In Table~\ref{tab:measures}, the portfolio of measures has been ordered according to the timescales required for their implementation. Various measures can be launched by us immediately or realized within a medium time period. The third category requires our strategic preparations, which involve coordinated efforts with additional stakeholders.

Overall, the commitment and joint effort of all stakeholders -- funding agencies, institutional bodies, computing centers, science managers, faculty, scientists, students -- are needed to master the challenges of sustainability in digital transformation in Universe and Matter research.

\backmatter

\bmhead{Supplementary information}
Data Availability Statement:
No explicit data were used in this study.

\bmhead{Acknowledgments}
This work is supported by the Ministry of Innovation, Science, and Research of the State of North Rhine-Westphalia, and by the Federal Ministry of Education and Research (BMBF) in Germany through the ErUM-Data-Hub project 05D21PA1. We acknowledge the useful support of the DeepL translator.

\begin{appendices}
\section{Twelve Guiding Questions}
\label{app:questions}
The workshop on Sustainability in the Digital Transformation of Basic Research on the Universe and Matter was structured around twelve prepared questions, listed below \cite{sustainability_workshop2023}.

\paragraph{Hardware \& Research Data}
\begin{enumerate}
\item
Footprint: Constructing a comprehensive picture of the footprint of all ErUM-Data related activities. Where does quantitative knowledge exist, where is it lacking? What resource needs do you see, what opportunities for savings? What innovations are needed to keep sustainable use of resources in balance with demands? To what extent does continuing education play a role? How can feedback reduce a footprint through machine learning methods? 
\item
(Dynamic) Energy Supply: Where to locate \& operate computing systems incl. storage? How could a dynamic energy supply look like, which largely covers the needs of ErUM-Data related activities with renewable energies? What information flows would be required for this? What mechanisms and what dynamics are required on a supra-regional basis to create compensation possibilities for windless/sunless periods?
\item
Hardware Lifetime: How could prolonged / optimized usage of hardware resources in view of technology evolution be modeled beyond their usual lifetimes? What short- and medium-term monitoring would be required to signal indispensable replacements on the one hand, and to execute computing jobs matching their algorithmic requirements on prolonged or current hardware on the other?
\item
Hardware \& Algorithms: Which adaptive measures for hardware and algorithms could have a decisive impact on ErUM-Data? Which types of hardware (including e.g., GPU, TPU, FPGA, neuromorphic computing) could be considered and which automated mechanisms exist for adapting algorithms to non-specific or dedicated hardware? 
\item
Smart Data: Deciding when and how to discard information without losing scientific value, based on learning from nature and experiment. What mechanisms for transforming data to smart data can be envisioned, and how can evaluation and control of information gain or loss be accomplished? How can archiving and retrieving data be managed? 
\item
Cultural Change: What could a comprehensive educational area for rethinking, among other things, the use of computer hardware, actually required information (smart data), preparation of data packages (event loops versus event chunks), etc. look like? How can we change to a culture of data reuse? Assessment of ethical implications and risk assessment.
\end{enumerate}

\paragraph{Algorithms \& Mindset}
\begin{enumerate}
\setcounter{enumi}{6}
\item
Autonomization: We witness the transformation from the era of automation to an era of autonomization (e.g., unsupervised learning). Where will ErUM-Data benefit from autonomization, which innovations are necessary and how can the reliability of the autonomously obtained results be ensured?
\item
Inquiries \& Dynamics: How can input questions be posed to generate the best possible output from the machines? What relevance will dynamic learning algorithms and machines have for the field of ErUM-Data? 
\item
Algorithmics \& Software: Our thinking in algorithms and software has a direct impact on resource requirements. What can sustainable algorithm \& software engineering and an associated educational program in algorithm \& software development look like to get ErUM-Data to the forefront of developers?
\item
Machine Models: Pre-trained and generative models have a high potential for energy savings in both their creation and usage of machine learning. What innovations are needed to achieve a reliable routine operation?
\item
Injected Intelligence: How can reasoning by the physicist, mathematician, or any other kind of intelligence speed up the processes of learning or make them more energy efficient? What measures can we apply to avoid constantly reinventing the wheel? What can knowledge discovery of work already performed look like?
\item
Workflow \& Stakeholders: How can well-defined, reproducible workflows with high user dynamics (data analyses) be captured that remain functional in the long term? How can an overall picture be created with all stakeholders working together on a large-scale project for the benefit of sustainability across their departmental boundaries?
\end{enumerate}

\section{Energy consumption of data center components}
\label{app:breakdown}
For data centers as used in the Worldwide LHC Computing Grid WLCG, typical numbers for power consumption of modern computing hardware in 2023 are: 
\begin{itemize}
    \item \textbf{CPU} Modern CPUs have power consumption in the region $3$--$8$ W per core, e.g. dual AMD EPYC 7513 node with $480$ W for $64$ cores~\cite{Britton:isgc} and about $1700$ HS23 \href{https://w3.hepix.org/benchmarking/scores_HS23.html}{(HS23-table)}  or $280$ W / kHS23~\footnote{HS23 is standard CPU benchmark used in WLCG, see \href{https://w3.hepix.org/benchmarking}{(HS23)}.}.
    Extrapolated to the whole available WLCG computing capacity (used by the four LHC experiments) of about \textbf{$\boldsymbol{14}$ MHS23} this would correspond to \textbf{$\boldsymbol{3.8}$ MW} (in reality consumption is presumably larger since average CPU hardware is older).
    \item \textbf{Disk storage} Modern storage servers have a power consumption around $1$--$2$ W per TB, e.g. HP Raid-$6$ server with $14\times 16$ TB HDD and usable capacity of $192$ TB consumes $240$ W ($=1.2$ W/TB).  Extrapolated again to full WLCG disk capacity of \textbf{$\boldsymbol{870}$ PB} this would correspond to about \textbf{$\boldsymbol{1.1}$ MW}.
    \item \textbf{Tape storage} Estimates for power consumption for tape storage are typically factor $10$ lower than disk storage~\cite{Campana:chep23-wlcg-energy}, i.e. around $0.1$ W/TB.
    \item \textbf{Networking} Power consumption of network routers and services are rather small, typically at the level of $2$--$3\%$ for a WLCG data center     
\end{itemize}
There is an additional overhead for cooling ranging from 10\% for highly optimized sites with direct warm-water cooling to 40\% for traditional air-cooled systems. 

\end{appendices}

%\bibliography{sustainability}% common bib file

%\end{document}

\end{document}